\documentclass[%
 reprint,
 superscriptaddress,
 amsmath,amssymb,
 aps,
 prl,
 twocolumn,
]{revtex4}
\usepackage{graphicx}
\usepackage{dcolumn}
\usepackage{bm}
\usepackage{color}

\begin{document}

\title{Temporal precision of molecular events with regulation and feedback}

\author{Shivam Gupta}
\affiliation{Department of Physics and Astronomy, Purdue University, West Lafayette, IN 47907, USA}

\author{Sean Fancher}
\affiliation{Department of Physics and Astronomy, Purdue University, West Lafayette, IN 47907, USA}
\affiliation{Department of Physics and Astronomy, University of Pennsylvania, Philadelphia, PA 19104, USA}

\author{Hendrik C.\ Korswagen}
\affiliation{Hubrecht Institute, Royal Netherlands Academy of Arts and Sciences and University Medical Center Utrecht, 3584 CT Utrecht, Netherlands}

\author{Andrew Mugler}
\email{amugler@purdue.edu}
\affiliation{Department of Physics and Astronomy, Purdue University, West Lafayette, IN 47907, USA}

\begin{abstract}
Cellular behaviors such as migration, division, and differentiation rely on precise timing, and yet the molecular events that govern these behaviors are highly stochastic. We investigate regulatory strategies that decrease the timing noise of molecular events. Autoregulatory feedback increases noise. Yet, we find that in the presence of regulation by a second species, autoregulatory feedback decreases noise. To explain this finding, we develop a method to calculate the optimal regulation function that minimizes the timing noise. The method reveals that the combination of feedback and regulation minimizes noise by maximizing the number of molecular events that must happen in sequence before a threshold is crossed. We compute the optimal timing precision for all two-node networks with regulation and feedback,  derive a generic lower bound on timing noise, and discuss our results in the context of neuroblast migration during {\it Caenorhabditis elegans} development.
\end{abstract}

\maketitle

Precise timing is crucial for many biological processes including cell division \citep{bean2006coherence, nachman2007dissecting, schneider2004growth}, cell differentiation \cite{carniol2004threshold}, cell migration \cite{mentink2014cell}, embryonic development \cite{meinhardt1982models, tufcea2015critical}, and cell death \cite{roux2015fractional}. Ultimately the timing of these processes is governed by the timing of molecular events inside the cell. However, these events are inherently stochastic. Cells use regulatory networks to reduce this stochasticity, but the effects of particular regulatory features on timing precision remain poorly understood. We recently demonstrated that the time at which an accumulating molecular species crosses an abundance threshold is more precise if that species is regulated by a second species with its own stochastic dynamics \cite{gupta2018temporal}. In contrast, it was recently demonstrated that if the species is instead regulated by itself (feedback), then the crossing time is less precise \cite{ghusinga2017first}. Yet, feedback is common in many important timing processes. In yeast, the cyclin proteins that  cross an abundance threshold to initiate the cell cycle \cite{schneider2004growth} are subject to positive feedback \cite{dirick1991positive, cross1991potential, bean2006coherence}. In {\it Caenorhabditis elegans}, the mig-1 protein that crosses an abundance threshold to terminate migration in QR neuroblasts \cite{mentink2014cell} has been found in experiments on the sister QL lineage to be subject to feedback via Wnt signaling \cite{ji2013feedback}. This raises the question of why feedback is observed in key timing processes if it has been shown to decrease timing precision.

Here we investigate the combined effect of regulation and feedback on timing precision. We develop a gradient-descent approach to find the globally optimal regulation function for a given network topology that minimizes the timing noise. We find that, despite the fact that feedback generically increases timing noise when it acts alone, feedback decreases timing noise when it acts in combination with regulation by an external species. We explain the mechanisms behind this counterintuitive result, derive a generic lower bound on the timing noise, and discuss the relevance of our results to the timing of neuroblast migration in {\it C.\ elegans}.

\begin{figure}
\includegraphics[width=.5\textwidth]{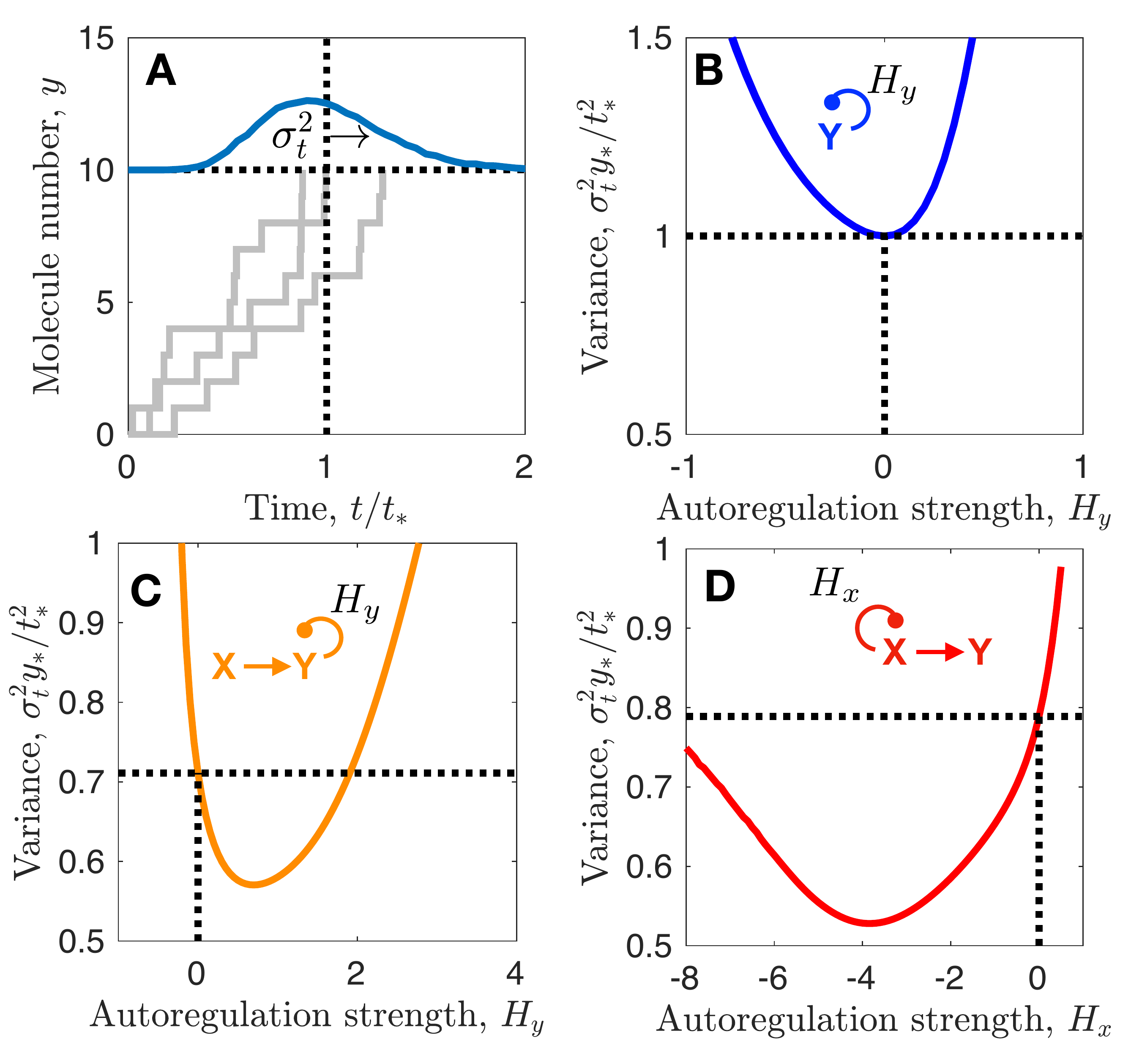}
\caption{\label{fig:fig1} Feedback increases timing precision in the presence but not absence of regulation. (A) A species $Y$ crosses a molecule-number threshold $y_*$ at mean time $t_*$ with timing variance $\sigma_t^2$. (B) Feedback increases the variance. However, in the presence of regulation by a second species $X$, feedback on either (C) $Y$ or (D) $X$ can decrease the variance. Parameters are $K_y = 2.5$ in B; $\alpha_0t_* = 10$, $H_x = -0.5$, $H_{xy} = -H_y$, $K_x = 15$, $K_y = 5$, and $K_{xy} = 6$ in C; $\alpha_0t_* = 10$, $H_y = 4$, $K_x = 10$, and $K_y = 7.5$ in D; and $y_* = 10$ throughout.}
\end{figure}

Consider a molecular species $Y$ that is produced over time and first reaches a molecule-number threshold $y_*$ at a particular time $t_*$ on average (Fig.\ \ref{fig:fig1}A). Stochasticity in the accumulation process leads to variability in the crossing time $t$. The timing noise is given by the variance $\sigma_t^2$. For unregulated production of $Y$, the time between each production event is exponentially distributed with mean $t_*/y_*$ and variance $(t_*/y_*)^2$. Because the production events are independent, the variances add, giving a total variance of $\sigma_t^2 = y_*(t_*/y_*)^2 = t_*^2/y_*$. Therefore we focus on the scaled variance $\sigma_t^2y_*/t_*^2$, whose value is $1$ for unregulated production.

First we investigate the effect of feedback on timing precision using a simple example: we suppose that the production rate of $Y$ is not a constant but rather is a simple sigmoidal function of the current number of molecules $y$,
\begin{equation}
\label{eqn:tanhy}
\beta(y) = \beta_0\{1+\tanh[H_y(y/K_y-1)]\},
\end{equation}
where positive (negative) $H_y$ corresponds to positive (negative) feedback, $|H_y|$ is the maximum steepness, $K_y$ is the molecule number at which $\beta$ is half-maximal, and $\beta_0$ is set to ensure that the average time at which $y$ first reaches $y_*$ is $t_*$. We calculate the variance $\sigma_t^2$ from the master equation by matrix inversion \cite{gupta2018temporal}. In Fig.\ \ref{fig:fig1}B we see that when there is no feedback ($H_y = 0$), the variance satisfies $\sigma_t^2y_*/t_*^2 = 1$, and that either positive or negative feedback increases the variance. This result is consistent with previous findings for a species that does not degrade \cite{ghusinga2017first}, and it has an intuitive explanation: a sequence of time-ordered stochastic events is most precisely timed if the mean time for each event to occur is equal, but feedback makes these times unequal.

Next we investigate the interplay of feedback and regulation by introducing a second species $X$ that is produced at a constant rate $\alpha_0$. The $Y$ production rate $\beta(x,y)$ is now a function of both molecule numbers $x$ and $y$. We find that if it is a simple sum $\beta(x,y) = f_1(x) + f_2(y)$ or product $\beta(x,y) = f_1(x)f_2(y)$ then feedback continues to generically increase the timing variance, but if we include a coupling term $\beta(x,y) = f_1(x)f_2(y)f_3(xy)$ the situation is different. Specifically, Fig.\ \ref{fig:fig1}C shows the case where
\begin{align}
\beta(x,y) = &\ \beta_0\{1+\tanh[H_x(x/K_x-1)]\} \nonumber \\
	&\times\{1+\tanh[H_y(y/K_y-1)]\} \nonumber \\
\label{eqn:tanhxy}
	&\times\{1+\tanh[H_{xy}(xy/K_{xy}^2-1)]\}.
\end{align}
We see that with no feedback ($H_y = 0$) we have $\sigma_t^2y_*/t_*^2 < 1$, which demonstrates that regulation by a second species increases the timing precision as found previously \cite{gupta2018temporal}. However, now we also see that with positive feedback ($H_y>0$), the variance can be even lower. Together with Fig.\ \ref{fig:fig1}B, this result implies that although feedback increases timing noise in the absence of regulation, it can decrease timing noise in the presence of regulation.

Similarly we investigate the case where the feedback occurs on $X$, not $Y$. We take the production rates of $x$ and $y$ to be
\begin{align}
\label{eqn:tanhx}
\alpha(x) =\ &\alpha_0\{1+\tanh[H_x(x/K_x-1)]\}, \\
\beta(x) =\ &\beta_0\{1+\tanh[H_y(x/K_y-1)]\},
\end{align}
respectively. We see in Fig.\ \ref{fig:fig1}D that with negative feedback ($H_x<0$) the variance is lower than with no feedback ($H_x = 0$), again implying that feedback can reduce timing noise when coupled to regulation.

To understand this effect, we develop a gradient-descent method to find the optimal regulation that minimizes the timing variance. The regulation is specified by the $X$ and $Y$ production rates $\alpha(x,y)$ and $\beta(x,y)$, respectively, which each depend on the molecule numbers $x$ and $y$ in general, but whose dependencies will later be restricted to consider particular feedback topologies. The probability of first reaching $y=y_*$ at time $t$ is $P(t) = \sum_{\{\vec{s}\}}P(t|\vec{s})P(\vec{s})$, where
\begin{align}
\label{eq:Ps}
P(\vec{s}) =\ &\prod_{i=0} ^{S-1} \frac{r_i}{k_i}, \\
\label{eq:Pts}
P(t|\vec{s}) =\ &\left(\prod_{i=0}^{S-1}\int_0^\infty dt_ik_ie^{-k_it_i}\right)\delta\left(t-\sum_{j=0}^{S-1}t_j\right).
\end{align}
In Eq.\ \ref{eq:Ps}, $P(\vec{s})$ is the probability of taking a path $\vec{s}$ from $(x_0,y_0) = (0,0)$ to $(x_S, y_S) = (x_S, y_*)$ for any nonnegative $x_S$, where $S$ is the length of the path. Each step $i$ takes the system out of state $(x_i,y_i)$ with rate $k_i = \alpha(x_i,y_i) + \beta(x_i,y_i)$ and into a new state with probability $r_i/k_i$, where the new state is either $(x_i+1,y_i)$ with $r_i = \alpha(x_i,y_i)$ or $(x_i,y_i+1)$ with $r_i = \beta(x_i,y_i)$. In Eq.\ \ref{eq:Pts}, $P(t|\vec{s})$ is the probability that traversing the given path $\vec{s}$ takes a time $t$. The first term integrates over all values of each step's transition time $t_i$, which is exponentially distributed with rate $k_i$, and the second term ensures that the sum of these transition times is $t$.
From $P(t)$ we calculate the moments \cite{supp}, of which the first two are
\begin{align}
\label{eqn:meant}
\langle t \rangle =\ &\sum_{\{\vec{s}\}}P(\vec{s})\sum_{i=0}^{S-1}\frac{1}{k_i},\\
\label{eqn:meant2}
\langle t^2 \rangle =\ &\sum_{\{\vec{s}\}}P(\vec{s})\left[\left(\sum_{i=0}^{S-1}\frac{1}{k_i^2}\right)+\left(\sum_{j=0}^{S-1}\frac{1}{k_j}\right)^2\right].
\end{align}
The optimal regulation function minimizes $\langle t^2 \rangle$ at fixed $\langle t \rangle = t_*$. Therefore, defining a vector $\vec{\gamma}$ whose components are all components of both the $\alpha(x,y)$ and $\beta(x,y)$ matrices, we initialize $\vec{\gamma}$ to satisfy $\langle t \rangle = t_*$ and update it as
\begin{equation}
\label{eqn:update}
\vec{\gamma}^{(n+1)} = \vec{\gamma}^{(n)} - \epsilon \vec{u}.
\end{equation}
Here $\epsilon \ll 1$, and $\vec{u}$ is such that $\vec{u}\cdot\nabla _\gamma \langle t^2\rangle$ is maximized with respect to the constraints $\vec{u}\cdot\nabla _\gamma \langle t\rangle = 0$ and $|u|^2=1$.

\begin{figure}
\includegraphics[width=.5\textwidth]{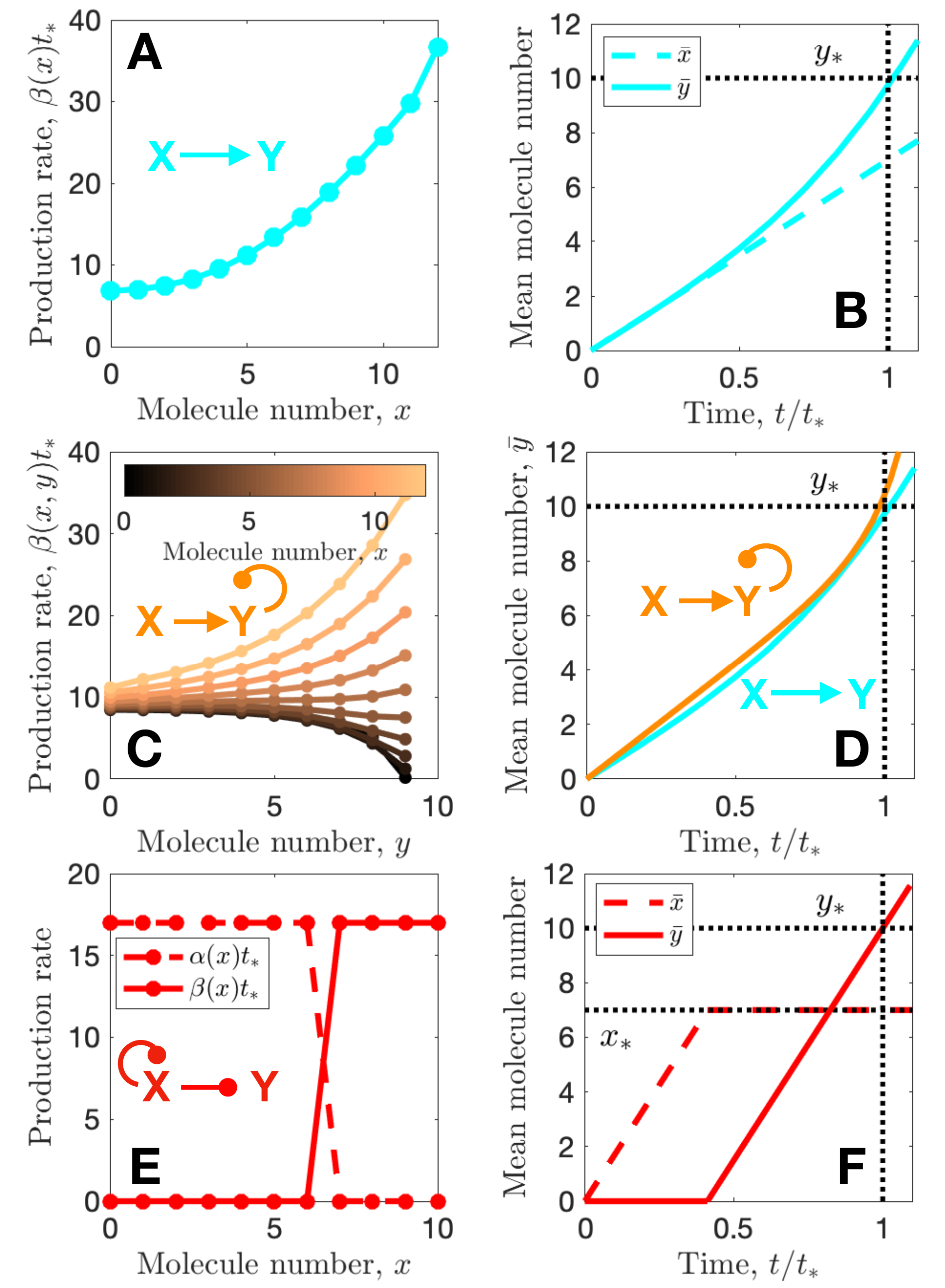}
\caption{\label{fig:fig2} Optimal regulation functions that minimize timing variance. (A) Without feedback, $X$ activates $Y$, (B) allowing $\bar{y}$ to accelerate before crossing $y_*$. (C) With feedback on $Y$, $X$ acts as a ``timer'' for $Y$, allowing $Y$ to self-repress at early times and self-activate at late times, and (D) providing further, late-phase acceleration of $\bar{y}$. (E) With feedback on $X$, it represses itself and activates $Y$ sharply, (F) resulting in kinked dynamics where $\bar{x}$ and $\bar{y}$ growth are separated in time. Parameters are $\alpha_0t_* = 7$ (A-D), $x_* = 7$ (E, F), and $y_* = 10$ throughout.}
\end{figure}

First we apply this method to the case where $X$ regulates $Y$ with no feedback. Thus, we fix $\alpha = \alpha_0$ and optimize $\beta(x)$. Figure \ref{fig:fig2}A shows the result, and we see that the optimal $\beta(x)$ is an increasing function of $x$ (i.e., $X$ activates $Y$). The reason, clear from the mean dynamics in \ref{fig:fig2}B, is that as $x$ increases over time, $\beta(x)$ increases over time, which causes $y$ to accelerate. The acceleration allows $\bar{y}$ to cross $y_*$ with a large slope, reducing the uncertainty of the crossing time. We observed this effect previously with Hill-function activation \cite{gupta2018temporal}, but the optimal regulation function was unknown.

Next we keep $\alpha = \alpha_0$, but we allow feedback on $Y$ and find the optimal $\beta(x,y)$. Figure \ref{fig:fig2}C shows the result, and we see that the optimal $\beta(x,y)$ depends on $y$, confirming that feedback is beneficial in the presence of regulation. Specifically, we see that $\beta(x,y)$ decreases with $y$ (negative feedback) when $x$ is small, and increases with $y$ (positive feedback) when $x$ is large. These two properties are also exhibited by Eq.\ \ref{eqn:tanhxy} with $H_x < 0$, $H_y > 0$, and $H_{xy} < 0$ as in Fig.\ \ref{fig:fig1}C. The first property ensures that $Y$ is not prematurely activated at early times when $x$ is small. The second property provides an additional acceleration of $y$ at late times when $x$ is large. Thus, $X$ acts as a ``timer'' for $Y$, allowing $Y$ to apply self-amplification only at late times. This has two advantages, as seen in Fig.\ \ref{fig:fig2}D: (i) it increases the slope of $\bar{y}$ at crossing, beyond that without feedback; and (ii) it allows the acceleration to begin at a $\bar{y}$ value that is already close to $y_*$, thus reducing trajectory-to-trajectory variability caused by prolonged self-amplification \cite{ghusinga2017first}.

Finally we consider the case where feedback acts on $X$ instead of $Y$. Here, to provide a reasonable constraint on $x(t)$, we introduce a bound $x_*$ and restrict $\alpha(x)$ such that $\bar{x}(t) \le x_*$ over the range $0 \le t \le t_*$. The optimal regulation functions $\alpha(x)$ and $\beta(x)$ are shown in Fig.\ \ref{fig:fig2}E. We see that $X$ represses itself and activates $Y$, and that both regulation functions have a sharp transition when $x = x_*$. We see in Fig.\ \ref{fig:fig2}F that the resulting dynamics are sharply kinked.

To understand the sharp nature of the optimal solution in Fig.\ \ref{fig:fig2}E and F, we investigate our optimization scheme (Eqs.\ \ref{eq:Ps}-\ref{eqn:update}) analytically. The analytic version of Eq.\ \ref{eqn:update} is $0 = \gamma_i\partial_{\gamma_i}(\langle t^2\rangle - \lambda\langle t\rangle)$, where the Lagrange multiplier $\lambda$ enforces $\langle t \rangle = t_*$, and the factor of $\gamma_i$ in front enforces $\gamma_i > 0$ \cite{supp}. By inserting Eqs.\ \ref{eqn:meant} and \ref{eqn:meant2} into this condition, we show \cite{supp} that it is satisfied when (i) $\alpha$ and $\beta$ are such that all possible paths $\vec{s}$ to reach $y = y_*$ have the same length $S$, and (ii) all transition rates along each of these paths are equal. Each such set of equal-length, constant-velocity paths is a local optimum, and the global optimum that minimizes the timing variance is the set for which (iii) the path length $S$ is as large as possible. More generally, if only property (ii) is satisfied, we show \cite{supp} that the timing variance satisfies
\begin{equation}
\label{eqn:variance_path}
\frac{\sigma_t^2}{t_* ^2} = \frac{\sigma_S^2}{\langle S \rangle ^2} + \frac{1}{\langle S \rangle},
\end{equation}
where $\langle S\rangle$ and $\sigma_S^2$ are the mean and variance of the path lengths, weighted by the path probabilities $P(\vec{s})$. Clearly the variance is minimized when $\sigma_S^2 = 0$ and $\langle S\rangle$ is as large as possible, consistent with properties (i) and (iii) above, respectively.

Now we can understand why the the optimal solution in Fig.\ \ref{fig:fig2}E and F looks the way it does. The sharp nature of the regulation functions ensures that at early times only $x$ changes, and at late times only $y$ changes, confining the stochastic dynamics to only one possible path in ($x$, $y$) space [property (i)]. The values of $\alpha$ and $\beta$, when they are nonzero, are constant and equal to each other, ensuring that the velocity along this path is constant [property (ii)]. Finally, both $x$ and $y$ attain their maximal values $x_*$ and $y_*$, ensuring that the path is as long as possible [property (iii)].

\begin{figure}
\includegraphics[width=.5\textwidth]{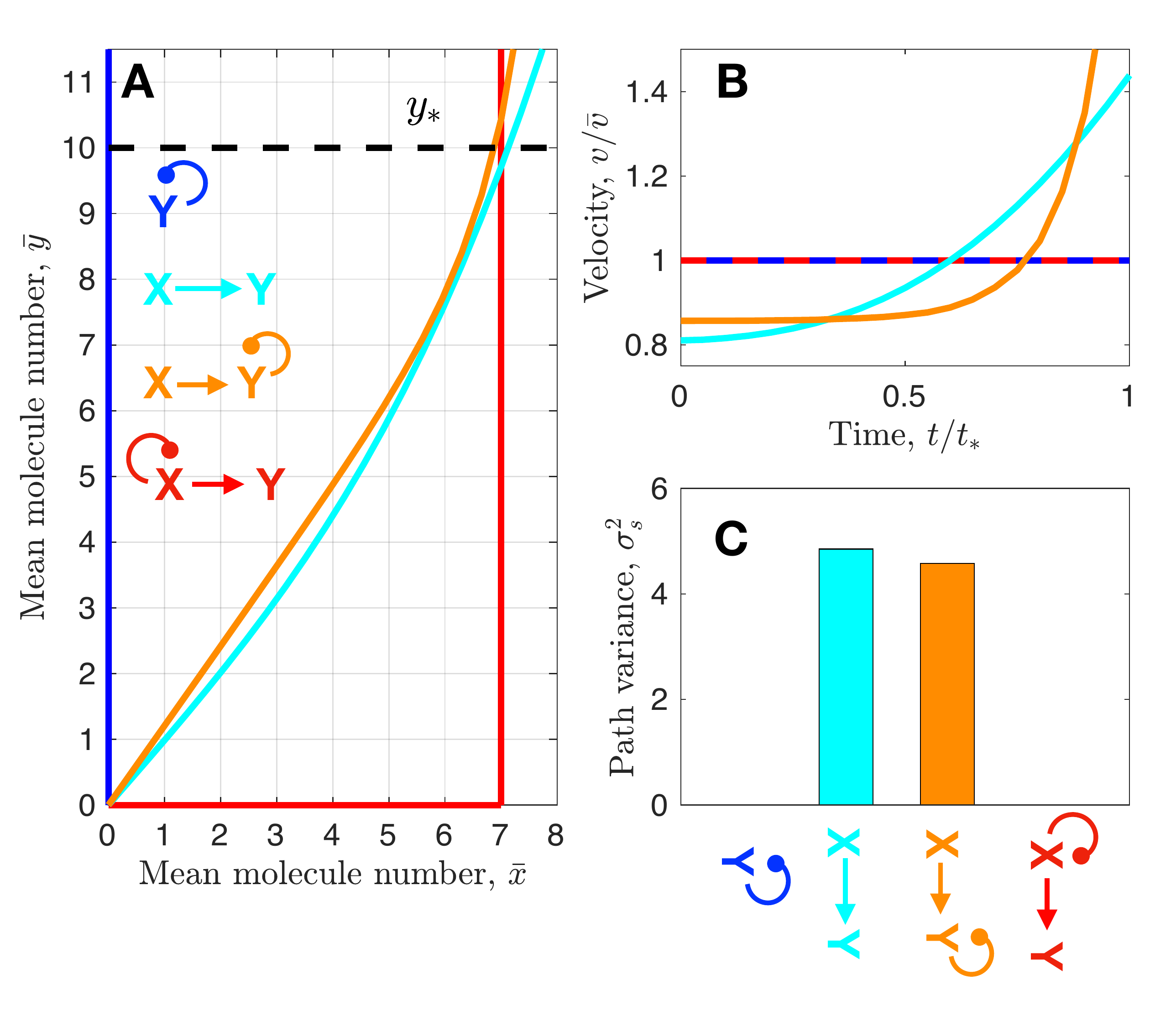}
\caption{\label{fig:fig3} Properties that minimize timing variance: (A) large path length $S$, (B) constant velocity $v(t)$ along path, and (C) small path length variance $\sigma^2_S$. Parameters as in Fig.\ \ref{fig:fig2}.}
\end{figure}

Indeed, Fig.\ \ref{fig:fig3} shows the optimal solutions for all of the networks considered thus far in terms of these three properties. Specifically, Fig.\ \ref{fig:fig3}A shows the mean dynamics in ($x$, $y$) space; Fig.\ \ref{fig:fig3}B shows the velocity $v(t) = \sqrt{(d\bar{x}/dt)^2+(d\bar{y}/dt)^2}$ along this path, normalized by its time average $\bar{v} = t_*^{-1}\int_0^{t_*}dt\ v(t)$; and Fig.\ \ref{fig:fig3}C shows the variance $\sigma_S^2$ in the path length across all paths. With only $Y$ and no $X$ (blue), there is only one possible path (Fig.\ \ref{fig:fig3}A), and therefore $\sigma_S^2 = 0$ (Fig.\ \ref{fig:fig3}C). The optimal solution has constant velocity along the path (Fig.\ \ref{fig:fig3}B), which is achieved with no feedback. When $X$ regulates $Y$ (cyan, orange), the mean path extends into the ($x$, $y$) plane (Fig.\ \ref{fig:fig3}A), which increases its length and thus lowers the timing variance. However, it also makes the velocity non-constant (Fig.\ \ref{fig:fig3}B) and allows for many possible paths such that $\sigma_S^2 > 0$ (Fig.\ \ref{fig:fig3}C). Only upon allowing $X$ to also regulate itself (red) does the path become as long as possible (Fig.\ \ref{fig:fig3}A), constant-velocity (Fig.\ \ref{fig:fig3}B), and unique (Fig.\ \ref{fig:fig3}C).

\begin{figure}
\includegraphics[width=.5\textwidth]{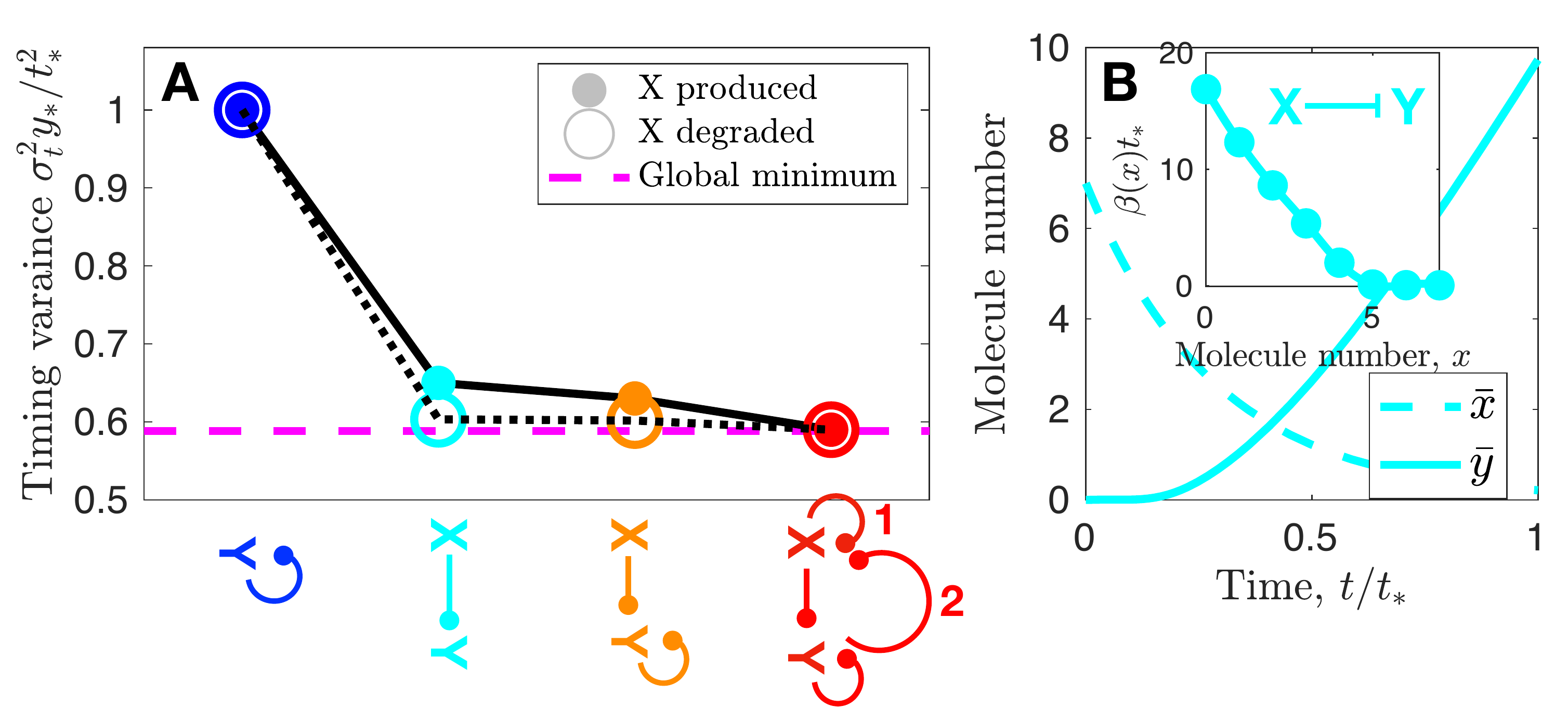}
\caption{\label{fig:fig4} (A) Ranking of timing variance for all one- and two-node networks. Global minimum is $\sigma_t^2y_*/t_*^2 = y_*/(x_*+y_*)$. In red network, link 1, 2, or both is required. Parameters as in Fig.\ \ref{fig:fig2}. (B) Mean dynamics and regulation function (inset) for case when $X$ is degraded. Here $\alpha_0t_* =  3.5$.}
\end{figure}

The minimal values of the timing variance for the networks are shown by the filled circles in Fig.\ \ref{fig:fig4}A. We see that the single species $Y$ achieves the standard $\sigma_t^2y_*/t_*^2 = 1$ (blue), regulation by $X$ lowers the variance (cyan), feedback on $Y$ lowers it further (orange), and regulation of $X$ lowers it to the global minimum given by Eq.\ \ref{eqn:variance_path} with $\sigma_S^2 = 0$ and $\langle S\rangle = x_* + y_*$, namely $\sigma_t^2y_*/t_*^2 = y_*/(x_*+y_*)$. Because the results in Fig.\ \ref{fig:fig4}A are minima, it does not matter in the last case whether the regulation of $X$ is by $X$ itself (red link 1), by $Y$ (red link 2), or both; the optimal regulation functions will produce the red path in Fig.\ \ref{fig:fig3} regardless.

Thus far we have only considered the scenario where $X$ is produced over time. However, $X$ could alternatively be degraded over time \cite{gupta2018temporal}. In the cases where $X$ is unregulated (cyan, orange), this corresponds to replacing its production propensity $\alpha_0$ (for $x\to x+1$) with a degradation propensity $\alpha_0 x$ (for $x\to x-1$). The resulting minimal values of the timing variance are shown by the open circles in Fig.\ \ref{fig:fig4}A, and we see that they are lower than the corresponding values when $X$ is produced over time (filled circles). The reason, illustrated for the case where $X$ regulates $Y$ in Fig.\ \ref{fig:fig4}B, is that when $X$ is produced over time it increases linearly (Fig.\ \ref{fig:fig2}B dashed), whereas when $X$ is degraded over time it decreases exponentially (Fig.\ \ref{fig:fig4}B dashed). The curvature of the exponential begins to approximate the kinked dynamics of the globally optimal solution (Fig.\ \ref{fig:fig2}F dashed). Specifically, $X$ is most dynamic at early times (Fig.\ \ref{fig:fig4}B dashed), and $Y$ is only produced once $x$ drops below a particular value (Fig.\ \ref{fig:fig4}B inset) allowing it to be most dynamic at late times (Fig.\ \ref{fig:fig4}B solid). Thus, even without feedback, the nonlinear dynamics of a degraded regulator allow its target to more closely approach the globally optimal timing precision.

How can these results be tested in experimental systems? Our findings suggest that a cellular process where timing precision is important should be governed by a molecular network with both multistep regulation and feedback, particularly one in which every species is subject to regulation as in Fig.\ \ref{fig:fig4}A (red). An experimental example in which timing precision is particularly well studied is neuroblast migration in developing {\it C.\ elegans} larvae. Here, the QR neuroblast produces a protein called mig-1 that crosses an abundance threshold to terminate migration; overproduction causes undermigration and vice versa \cite{mentink2014cell}. It was recently discovered in the sister QL lineage that mig-1 is subject to both regulation and negative feedback via canonical Wnt signaling \cite{ji2013feedback}. Specifically, mig-1 activates one or more Wnt signaling factors, which in turn repress mig-1. These interactions form a network of the red type in Fig.\ \ref{fig:fig4}A (with link 2), where $X$ is the Wnt factor and $Y$ is mig-1, which is precisely the class of networks that we predict achieve the globally minimum timing noise. We anticipate that other biological processes where timing precision is paramount will be governed by interaction networks in this class.

We have developed a gradient-descent approach that provides the optimal regulation functions for a given network topology that minimize the timing noise of a threshold-crossing event. The approach has revealed that feedback reduces timing noise in the presence but not absence of regulation because the combination of the two increases the number of transitions that must happen sequentially in molecular state space. More generally, our work suggests a perspective where noise is not minimized by finding the right network topology, but rather by finding the right combination of regulation functions that produce a path through state space that is as long, steady, and unique as possible. Our approach is straightforward to generalize to larger and more complex networks, and we anticipate that this perspective applies broadly to biological processes where timing is crucial.

\acknowledgments
This work was supported by Human Frontier Science Program grant RGP0030/2016 and Simons Foundation grant 376198.

\onecolumngrid
\section{Supplementary Material}

\appendix
\section{Calculation of the moments of the first passage time}

Using Eqs.\ 5 and 6 of the main text, we write the first passage time distribution as
\begin{align}
P(t) &= \sum_{\{\vec{s}\}}P(\vec{s})P(t|\vec{s}) \nonumber\\
&= \sum_{\left\{\vec{s}\right\}}\left(\prod_{i=0}^{S-1}\frac{r_{i}}{k_{i}}\right)\left(\prod_{j=0}^{S-1}\int_{0}^{\infty}dt_{j}k_{j}e^{-k_{j}t_{j}}\right)\delta\left(t-\sum_{\ell=0}^{S-1}t_{\ell}\right) \nonumber\\
&= \sum_{\left\{\vec{s}\right\}}\left(\prod_{i=0}^{S-1}\int_{0}^{\infty}dt_{i}r_{i}e^{-k_{i}t_{i}}\right)\delta\left(t-\sum_{j=0}^{S-1}t_{j}\right).
\label{TFPTD}
\end{align}
The $n$th moment is
\begin{align}
\left\langle t^{n}\right\rangle &= \int_{0}^{\infty}dt\ t^{n}P(t) \nonumber\\
&= \int_{0}^{\infty}dt\ t^{n}\sum_{\left\{\vec{s}\right\}}\left(\prod_{i=0}^{S-1}\int_{0}^{\infty}dt_{i}r_{i}e^{-k_{i}t_{i}}\right)\delta\left(t-\sum_{j=0}^{S-1}t_{j}\right) \nonumber\\
&= \sum_{\left\{\vec{s}\right\}}\left(\prod_{i=0}^{S-1}\int_{0}^{\infty}dt_{i}r_{i}e^{-k_{i}t_{i}}\right)\left(\sum_{j=0}^{S-1}t_{j}\right)^{n}.
\label{tnexp}
\end{align}
Specifically, the first and second moments are
\begin{align}
\left\langle t\right\rangle &= \sum_{\left\{\vec{s}\right\}}\left(\prod_{i=0}^{S-1}\int_{0}^{\infty}dt_{i}r_{i}e^{-k_{i}t_{i}}\right)\left(\sum_{j=0}^{S-1}t_{j}\right) \nonumber\\
&= \sum_{\left\{\vec{s}\right\}}\left(\prod_{i=0}^{S-1}\frac{r_{i}}{k_{i}}\right)\sum_{j=0}^{S-1}\frac{1}{k_{j}}\nonumber\\
&= \sum_{\left\{\vec{s}\right\}}P(\vec{s})\sum_{j=0}^{S-1}\frac{1}{k_{j}}
\label{FPTmom1}
\end{align}
and
\begin{align}
\left\langle t^{2}\right\rangle &= \sum_{\left\{\vec{s}\right\}}\left(\prod_{i=0}^{S-1}\int_{0}^{\infty}dt_{i}r_{i}e^{-k_{i}t_{i}}\right)\left(\sum_{j=0}^{S-1}t_{j}\right)^{2} \nonumber\\
&= \sum_{\left\{\vec{s}\right\}}\left(\prod_{i=0}^{S-1}\int_{0}^{\infty}dt_{i}r_{i}e^{-k_{i}t_{i}}\right)\left(\sum_{j=0}^{S-1}t_{j}^{2}+\sum_{j=0}^{S-2}\sum_{\ell=j+1}^{S-1}2t_{j}t_{\ell}\right) \nonumber\\
&= \sum_{\left\{\vec{s}\right\}}\left(\prod_{i=0}^{S-1}\frac{r_{i}}{k_{i}}\right)\sum_{j=0}^{S-1}\sum_{\ell=j}^{S-1}\frac{2}{k_{j}k_{\ell}} \nonumber\\
&= \sum_{\left\{\vec{s}\right\}}\left(\prod_{i=0}^{S-1}\frac{r_{i}}{k_{i}}\right)\left(\left(\sum_{j=0}^{S-1}\frac{1}{k_{j}^{2}}\right)+\left(\sum_{j=0}^{S-1}\frac{1}{k_{j}}\right)^{2}\right),\nonumber\\
&= \sum_{\left\{\vec{s}\right\}}P(\vec{s})\left(\left(\sum_{j=0}^{S-1}\frac{1}{k_{j}^{2}}\right)+\left(\sum_{j=0}^{S-1}\frac{1}{k_{j}}\right)^{2}\right),
\label{FPTmom2}
\end{align}
as in Eqs.\ 7 and 8 of the main text, where the last line in each case recalls Eq.\ 5 from the main text.

\section{Analytic minimization of timing variance using Lagrange multipliers}

To find the minimum variance when the mean is fixed to be $t^{*}$, we utilize Lagrange multipliers. Because the variance is a function of only the first and second moments and is monotonically increasing with the second moment, finding the minimum of the variance with a fixed mean is equivalent to finding the minimum of the second moment with a fixed mean. Thus, the set of $r_{\ell}$ values which produces the minimum variance is the set which solves
\begin{equation}
0 = \frac{\partial}{\partial r_{\ell}}\left(\left\langle t^{2}\right\rangle-\lambda\left\langle t\right\rangle\right)
\label{lagrangeS}
\end{equation}
for Lagrange multiplier $\lambda$.

However, Eq.\ \ref{lagrangeS} raises an issue. Assume that $x_*=y_*=1$. In this case, there are only three possible rates $\alpha_{xy}$ and $\beta_{xy}$, namely $\alpha_{00}$, $\beta_{00}$, and $\beta_{10}$. There are also only two possible paths: $\vec{s}_{1} = \left\lbrack\left\{0,0\right\},\left\{0,1\right\}\right\rbrack$ and $\vec{s}_{2} = \left\lbrack\left\{0,0\right\},\left\{1,0\right\},\left\{1,1\right\}\right\rbrack$. Putting these rates and paths into Eqs.\ \ref{FPTmom1} and \ref{FPTmom2} yields
\begin{align}
\left\langle t\right\rangle &= \frac{\beta_{00}}{\alpha_{00}+\beta_{00}}\frac{1}{\alpha_{00}+\beta_{00}}
	+\frac{\alpha_{00}}{\alpha_{00}+\beta_{00}}\frac{\beta_{10}}{\beta_{10}}\left(\frac{1}{\alpha_{00}+\beta_{00}}+\frac{1}{\beta_{10}}\right) \nonumber\\
&= \frac{1}{\alpha_{00}+\beta_{00}}\left(1+\frac{\alpha_{00}}{\beta_{10}}\right)
\label{1b1mom1}
\end{align}
and
\begin{align}
\left\langle t^{2}\right\rangle &= \frac{\beta_{00}}{\alpha_{00}+\beta_{00}}\frac{2}{\left(\alpha_{00}+\beta_{00}\right)^{2}}+\frac{\alpha_{00}}{\alpha_{00}+\beta_{00}}\frac{\beta_{10}}{\beta_{10}}\left(\frac{2}{\left(\alpha_{00}+\beta_{00}\right)^{2}}+\frac{2}{\left(\alpha_{00}+\beta_{00}\right)\beta_{10}}+\frac{2}{\beta_{10}^{2}}\right) \nonumber\\
&= \frac{2}{\left(\alpha_{00}+\beta_{00}\right)^{2}}\left(1+\frac{\alpha_{00}}{\beta_{10}}+\frac{\alpha_{00}\left(\alpha_{00}+\beta_{00}\right)}{\beta_{10}^{2}}\right).
\label{1b1mom2}
\end{align}
By putting Eqs.\ \ref{1b1mom1} and \ref{1b1mom2} into Eq.\ \ref{lagrangeS} and solving the resulting system of equations, one obtains that some rates must be negative or even undefined depending on the order in which they are solved. Since negative rates are unphysical, we can enforce positivity by making the substitutions $\alpha_{xy}=\text{exp}\left(a_{xy}\right)/t^{*}$ and $\beta_{xy}=\text{exp}\left(b_{xy}\right)/t^{*}$ and finding the minimum variance in $\left(a_{xy},b_{xy}\right)$ space rather than $\left(\alpha_{xy},\beta_{xy}\right)$ space. This procedure can be done without ever leaving $\left(\alpha_{xy},\beta_{xy}\right)$ space by noting that $\partial/\partial a = \left(\partial\alpha/\partial a\right)\partial/\partial\alpha = \alpha\left(\partial/\partial\alpha\right)$ and similarly that $\partial/\partial b = \beta\left(\partial/\partial\beta\right)$. This allows Eq.\ \ref{lagrangeS} to be rewritten as
\begin{equation}
0 = r_{\ell}\frac{\partial}{\partial r_{\ell}}\left(\left\langle t^{2}\right\rangle-\lambda\left\langle t\right\rangle\right).
\label{lagrangepos}
\end{equation}
Putting Eqs.\ \ref{1b1mom1} and \ref{1b1mom2} into Eq.\ \ref{lagrangepos} yields two possible solutions to the resulting equations: $\left\lbrack\beta_{00},\alpha_{00},\beta_{10}\right\rbrack=\left\lbrack 1/t_{*},0,\beta_{10}\right\rbrack$ with $\sigma^{2}=t_*^{2}$ for any value of $\beta_{10}$ or $\left\lbrack\beta_{00},\alpha_{00},\beta_{10}\right\rbrack=\left\lbrack 0,2/t_{*},2/t_{*}\right\rbrack$ with $\sigma^{2}=t_*^{2}/2$. Of important note is the fact that when $\alpha_{00}=0$ only the $\vec{s}_{1}$ path is available, while when $\beta_{00}=0$ only the $\vec{s}_{2}$ path is available. Thus, the variance is seen to be extremized when only one possible path is available and all rates along that path are equal. Additionally, the longer path yields a smaller variance.

This can be seen to be a simple case of a larger trend. For any possible values of $x_{*}$ and $y_*$ it is possible to choose a set of reaction rates such that there is only one possible path through $\left(x,y\right)$ space. When this is done, the product terms in Eqs. \ref{FPTmom1} and \ref{FPTmom2} becomes identically 1 since $r_{i}=k_{i}$ must be true along the one possible path. All other paths will have $r_{i}=0$ for some $i$ and will thus not contribute. This allows Eq. \ref{lagrangepos} to be easily calculated for any $r_{\ell}$ that is in the single possible path,
\begin{align}
0 &= r_{\ell}\frac{\partial}{\partial r_{\ell}}\left(\left(\sum_{i=0}^{S-1}\frac{1}{r_{i}^{2}}\right)+\left(\sum_{i=0}^{S-1}\frac{1}{r_{i}}\right)^{2}-\lambda\left(\sum_{i=0}^{S-1}\frac{1}{r_{i}}\right)\right) \nonumber\\
&= \frac{\lambda}{r_{\ell}}-\frac{2}{r_{\ell}^{2}}-\frac{2}{r_{\ell}}\left(\sum_{i=0}^{S-1}\frac{1}{r_{i}}\right)
\label{lagrangepath}
\end{align}
Eq. \ref{lagrangepath} is true for all $r_{\ell}$ along the single path if and only if all $r_{\ell}$ along that path have the same value, which, from the restriction that the mean first passage time must be $t^{*}$ and Eq. \ref{FPTmom1}, means $r_{\ell}=S/t^{*}$. Putting these values back into Eq. \ref{FPTmom2} then allows the variance to be simply calculated to be $\sigma^{2}={t^{*}}^{2}/S$.

Eq. \ref{lagrangepos} must hold for all off-path reactions as well. This can be seen to be true by noting that for all other paths at least one $r_{i}$ must be 0 in the product term. If $\ell\ne i$ this fact is not changed and that path will still have 0 contribution. If $\ell=i$ then the $r_{\ell}$ in front of the derivative operator will still force that path to have 0 contribution since no $k_{i}$ can be 0. Similarly, if $r_{\ell}$ is not a reaction that occurs at any state along the one possible path then the derivative will cause it to vanish since the contribution from the one possible path does not depend on rates that exist in other states, while if $r_{\ell}$ is a 0 rate that exists at a state in the one possible path then the factor of $r_{\ell}$ in front of the derivative will cause the whole expression to vanish. Thus, choosing a set of reaction rates such that there is a single possible path and all rates along that path are equal is a solution to Eq. \ref{lagrangepos} for all $r_{\ell}$. Additionally, since $\sigma^{2}={t^{*}}^{2}/S$, the longer that path is the smaller the variance will be. We state this result more generally by establishing three rules which state that the variance in first passage time is minimized when:

\begin{enumerate}
\item Variability in the possible path taken is minimized
\item Rate at which the system moves through state space is as constant as possible
\item The path length through state space is maximized
\end{enumerate}

\section{Derivation of the lower bound on timing variance}
If all rates are the same, $k_{i}=k$, then Eqs.\ \ref{FPTmom1} and \ref{FPTmom2} become
\begin{equation}
\left\langle t\right\rangle = \sum_{\{\vec{s}\}}P\left(\vec{s}\right)\frac{S}{k} = \frac{\left\langle S\right\rangle}{k}
\label{aveteqk}
\end{equation}
and
\begin{equation}
\left\langle t^{2}\right\rangle = \sum_{\{\vec{s}\}}P\left(\vec{s}\right)\left(\frac{S}{k^{2}}+\frac{S^{2}}{k^{2}}\right) = \frac{\left\langle S\right\rangle}{k^{2}}+\frac{\left\langle S^{2}\right\rangle}{k^{2}}.
\label{avet2eqk}
\end{equation}
We then have
\begin{equation}
\frac{\sigma_{t}^{2}}{\left\langle t\right\rangle^{2}} = \frac{\left\langle t^{2}\right\rangle-\left\langle t\right\rangle^{2}}{\left\langle t\right\rangle^{2}} = \frac{k^{2}}{\left\langle S\right\rangle^{2}}\left(\frac{\left\langle S\right\rangle}{k^{2}}+\frac{\left\langle S^{2}\right\rangle}{k^{2}}-\frac{\left\langle S\right\rangle^{2}}{k^{2}}\right) = \frac{1}{\left\langle S\right\rangle}+\frac{\sigma_{S}^{2}}{\left\langle S\right\rangle^{2}},
\label{NSR}
\end{equation}
as in Eq.\ 10 of the main text.


\end{document}